\documentstyle[cas]{article}

\begin{document}
\vspace{0.4cm}
\begin{flushright}
CERN-TH-7039/93 \\
UND-HEP-93-BIG03\\
\end{flushright}
\vspace{1.0cm}
\begin{center} \Large
{WEAK DECAYS OF CHARM HADRONS:  THE NEXT LESSON ON QCD --
AND POSSIBLY MORE!\footnote {Invited lecture given at the
Third Workshop on the Tau-Charm Factory, Marbella, Spain, June 1993}}
\end{center}
\vspace{0.4cm}
\begin{center}{\Large I.I. Bigi}\\
Theoretical Physics Division, CERN,CH-1211, Geneva 23, Switzerland
\footnote {during the academic year 1993/94}
\\
and
\\
Dept. of Physics, University of Notre Dame du Lac,
Notre Dame, IN 46556, U.S.A.
\footnote {permanent address}
\\e-mail address: VXCERN::IBIGI, BIGI@UNDHEP
\end{center}
\vspace{1.4cm}
\begin{center}
{\Large Abstract}
\end{center}
Second-generation theoretical technologies -- heavy quark
expansions, QCD sum rules and QCD simulations on the lattice --
are arriving on the scene, allowing a treatment of charm
decays that is genuinely based on QCD. The availability of those
theoretical tools strengthens the case for the need for
comprehensive precision measurements of charm decays. It is
pointed out that the decays of charm mesons as well as of charm
baryons have to be systematically analysed to gain control
over theoretical uncertainties. A $\tau$-charm factory with
$E_{c.m.}\leq 5.5$ GeV is optimally suited for such a program.
It would lead to a deeper understanding of QCD that
would also prepare us better for exploiting the
discovery potential anticipated in beauty decays.
In addition it is quite conceivable (though certainly not
guaranteed) that there arise fundamental
surprises, such as non-canonical rare $D$ decays,
$D^0$-$\bar D^0$ oscillations and CP violation in charm decays
which would dramatically alter our perspective on
Nature's fundamental forces.

\vspace{2.4cm}
\noindent CERN-TH-7039/93\\
October 1993
\newpage

\section{INTRODUCTION}

An instructive global perspective on the forces controlling the weak
decays of charm
hadrons can be gained by considering their lifetime ratios
\cite {DANILO}:
$$\frac {\tau (D^+)}{\tau (D^0)}\simeq 2.50\pm 0.05\; ,\; \; \;
\frac {\tau (\Xi _c^+)}{\tau (\Xi _c^0)}\simeq 4.0\pm 1.5\eqno(1)$$
$$\frac {\tau (D_s)}{\tau (D^0)}\simeq 1.13\pm 0.05\; ,\; \; \;
\frac {\tau (\Xi _c^+)}{\tau (\Lambda _c)}\simeq 2.0\pm 0.7\eqno(2)$$
$$\frac {\tau (\Lambda _c)}{\tau (D^0)}\simeq 0.51\pm 0.05\; ,\; \; \;
\; \frac {\tau (D^+)}{\tau (\Xi _c^0)}\simeq 10\pm 2.5\eqno(3)$$
The {\em qualitative} pattern in these ratios can be readily
understood by realizing that destructive interference (and to a
lesser degree `Weak Annihilation') affects meson lifetimes
whereas baryon lifetime ratios are shaped by the interplay of
`Weak Scattering' and destructive as well as constructive
interference. Yet a pessimist looks at these numbers, realizes
that the non-perturbative corrections that distinguish the various
decay rates are large, that everything is rather involved, and
concludes that
no quantitative insight can be gained here. An optimist on
the other hand
observes that these non-perturbative
effects -- while certainly large --
are not of an overwhelming size: only the extreme ratio
between the longest- and the shortestlived charm hadron amounts to
an order of magnitude. Furthermore there are so many transitions
\footnote {In addition to lifetimes there are semileptonic
branching ratios and lepton energy spectra as will be discussed
later.} that a theoretical description that might be of
uncertain validity when applied to a
single decay class will be cross-checked
in many highly non-trivial ways. The optimist thus looks at
charm decays as a unique opportunity to be educated about the
workings of QCD in a novel environment -- and that is the
attitude that I am going to advocate here.

This optimism is fed by the timely emergence and increasing maturity
level of second-generation theoretical technologies applicable here,
namely

$\bullet$ treatments based on QCD sum rules;

$\bullet$ expansions of the transition amplitudes in powers of
$1/m_Q$, $m_Q$ being the heavy flavour quark mass;

$\bullet$ numerical simulations of QCD on a lattice.

These methods are to be seen as complementing each other
rather than competing against each other. To cite but one
example: lattice simulations are just now reaching a level
where charm decays can be tackled \cite {MART},
whereas a treatment of beauty decays
will remain beyond our reach still for some time to come;
$1/m_Q$ expansions on the other hand can be expected to be
well behaved for beauty decays since $\mu _{had}/m_b
\ll 1$, while in charm decays
terms that are formally of higher order
in $1/m_c$
might turn out to be numerically important.
A detailed study of charm decays then serves a dual purpose: (i) They
provide a common arena for probing different theoretical
technologies.\break
(ii) Theoretical predictions based on $1/m_Q$ expansions have to
be viewed a priori as not better than semi-quantitative in charm decays;
yet their
confrontation with detailed data will allow us to infer the weight
of different non-perturbative corrections in charm decays and
extrapolate them to beauty decays.

The remainder of this talk will be organized as follows: in Sect. 2 I
sketch a description of inclusive charm decays that is intrinsically
connected with QCD; in Sect. 3 I address exclusive non-leptonic two-body
decay modes; in Sect. 4 I briefly discuss rare charm
decays before presenting a summary in Sect. 5.

\section{INCLUSIVE HEAVY-FLAVOUR DECAYS}

The widths for the weak decays of heavy-flavour hadrons $H_Q$ into an
inclusive final state $f$ can be calculated in QCD as an expansion in
powers of $1/m_Q$ \cite {VS,BU}:
$$\Gamma (H_Q\rightarrow f)=\frac{G_F^2m_Q^5}{192\pi ^3}|KM|^2
[c_3(f)\langle H_Q|\bar QQ|H_Q\rangle +
c_5(f)\frac {\langle H_Q|\bar Qi\sigma \cdot GQ|H_Q\rangle }
{m_Q^2}+$$
$$+\sum _i c_6^{(i)}(f)
\frac {\langle H_Q|\bar Q\Gamma _iq\bar q\Gamma _iQ|H_Q\rangle }
{m_Q^3}
+{\cal O}(1/m_Q^4)]\eqno(4), $$
where the dimensionless coefficients $c_i(f)$ depend on the parton
level characteristics of $f$ and on the ratios of the final-state
quark masses to $m_Q$; $KM$ denotes the appropriate combination of
weak mixing angles.

It is through the expectation values of the operators appearing on
the right-hand side of eq. (4) that the dependence on the decaying
{\em hadron} and on non-perturbative forces in general enters. Since
these are on-shell matrix elements one sees that
$\Gamma (H_Q\rightarrow f)$ is indeed expanded into a power series
in $\mu _{had}/m_Q$, as stated before. One can immediately read off
an important qualitative result from eq. (4): there are $no$
corrections of order $1/m_Q$ to {\em total} rates [although they
emerge for lepton {\em spectra} \cite {BUV}].
The leading non-perturbative
corrections thus scale like $1/m_Q^2$, i.e. they fade away quickly with
increasing $m_Q$.

While at present one is unable to determine the size of
these matrix elements
from first principles, one can relate them to other observables.
The expectation value of the chromomagnetic operator
$\bar Qi\sigma \cdot GQ$ can be extracted from the measured
hyperfine splitting between the pseudoscalar and vector states:
$$<P_Q|\bar Qi\sigma \cdot GQ|P_Q>\simeq
\frac {3}{2}(M_{V_Q}^2-M_{P_Q}^2)  \eqno(5)$$
with $V_Q=D^*[B^*]$ and $P_Q=D[B]$ for $Q=c[b]$.
For the baryons $\Lambda _Q$, on the other hand, it vanishes:
$$<\Lambda _Q|\bar Qi\sigma \cdot GQ|\Lambda _Q>\simeq 0
\eqno(6)$$
The scalar
operator $\bar QQ$ can be expanded again into a series of inverse
powers of $m_Q$:
$$<H_Q|\bar QQ|H_Q>= 1-\frac {<({\vec p})^2>}{2m_Q^2}+
\frac {3}{8} \cdot \frac {M_{V_Q}^2-M_{P_Q}^2}{m_Q^2} +
{\cal O}(1/m_Q^3), \eqno(7)$$
where $<({\vec p})^2>/2m_Q\equiv
<H_Q|\bar Q(i{\vec D})^2Q|H_Q>/2m_Q$ denotes the expectation value
for the kinetic energy of the heavy quark $Q$ moving inside the
hadron $H_Q$ under the influence of the gluon background field.
The first term on the right-hand side of eq. (7) reproduces the
simple parton model result, i.e. the `spectator ansatz'
leading to universal lifetimes and semileptonic branching ratios
for all hadrons of a given flavour $Q$; the
first two terms then represent the mean value of the
Lorentz time dilatation factor $\sqrt (1-v^2)$ that slows down the
decay of the quark $Q$ in a moving frame. The numerical size of
$<({\vec p})^2>$ is not known yet. Results from two analyses based on
QCD sum rules exist, though \cite {NEUBERT};
lattice simulations of QCD will be able to
extract this quantity in the near future \footnote {Lattice calculations
will actually yield more directly $<H_c|\bar cc|H_c>$.}; measuring the
mass of $\Lambda _b$ to 10 MeV accuracy will enable us to determine
the difference of $<({\vec p})^2>$ for the heavy-flavour meson and
baryon states.

The last term in eq. (4) incorporates the conventional non-spectator
effects, namely `Pauli Interference' and
`Weak Annihilation/Scattering'
and thus generates lifetime differences between the
different hadrons of a given heavy flavour.

The expansion parameter $\mu _{had}/m_b\sim 1/10$
for beauty decays is reasonably
small and one can expect the first few terms in the expansion to yield
a good approximation to the exact result; this also means that
non-perturbative corrections are smallish in inclusive beauty decays.
For charm decays on the other hand one finds $\mu _{had}/m_c
\sim 0.3$. While this quantity is at least smaller than unity, it is
not small. Contributions that are formally of higher order in
$1/m_Q$ can then become numerically important. Therefore one
expects to make predictions for charm decays that can at best
claim semi-quantitative validity. Yet there is another -- and I think
more fruitful --
perspective on this situation: charm decays serve as Nature's
microscope, magnifying the non-perturbative corrections
affecting beauty decays by factors $(m_b/m_c)^2\sim 10$ and even
$(m_b/m_c)^3\sim 30$ \cite {SHIFMAN}!

This situation can be illustrated by one topical
example, namely the size of the semileptonic branching ratios.
To leading order in $1/m_c$ one finds of course the parton model
result $BR_{SL}(D)\sim 15\%$. The first non-perturbative correction
arises on the $1/m_c^2$ level and reduces it considerably, yielding
$BR_{SL}(D)\sim 10 \%$ for $D^0$, $D^+$ and $D_s$ mesons. To order
$1/m_c^3$ one finally obtains a large splitting, namely
$BR_{SL}(D^+)\sim 16 \%$ and $BR_{SL}(D^0)\sim 8 \%$. Clearly,
on this level of the analysis there are large numerical
uncertainties. Yet those can be brought under control once
other observables have been measured and compared with the
theoretical predictions, namely

$\bullet$ $BR_{SL}(D_s)$, $BR_{SL}(\Lambda _c)$ and
$BR_{SL}(\Xi _c)$;

$\bullet$ the lepton spectra in the semileptonic decays of $D^0$,
$D^+$, $D_s$, $\Lambda _c$ and $\Xi _c$.

For it is the same set of operators, namely $\bar cc$,
$\bar ci\sigma \cdot Gc$ and
$(\bar c\Gamma _iq)(\bar q\Gamma _ic)$, that control all
these processes through order $1/m_c^3$.

\section{EXCLUSIVE NON-LEPTONIC TWO-BODY DECAYS}

\subsection{Phenomenological Models}

Phenomenological descriptions of two-body modes for
heavy-flavour hadrons were pioneered by the
authors of ref. \cite {BSW}.
There are
three main ingredients in such models: (i) One assumes factorization, i.e.
one uses $<M_1M_2|J_{\mu}J_{\mu}|D>\;  \simeq \; <M_1|J_{\mu}|D>\cdot
\break
<M_2|J_{\mu}|0>$ to describe $D\rightarrow M_1M_2$.
(ii) One employs one's favourite hadronic wavefunctions to
compute $<M_1|J_{\mu}|D>$. (iii) All two-body modes are then
expressed in terms of two free fit parameters $a_1$ and $a_2$ with
$a_1$ controlling the so-called ``class I''
$D^0\rightarrow M_1^+M_2^-$ and
$a_2$ the ``class II'' $D^0\rightarrow M_1^0M_2^0$
transitions; both quantities
contribute coherently to the ``class III''
$D^+\rightarrow M_1^0M_2^+$ transitions.

With these two free parameters $a_1$ and $a_2$
(and some considerable degree of
``poetic license'' in invoking strong final state interactions),
one obtains a decent fit for the $D^0$ and $D^+$ modes.
One has to point out, though, that this success is considerably
helped by the merciful imprecision in some of the branching ratios
measured so far.

Very recently Heavy Quark Symmetry and Chiral Symmetry (for the
light quarks) have been incorporated into
these models\cite {GATTO}; thus they
have presumably reached their final level of maturity.
To pass a final judgement on these phenomenological
models requires an analysis

$\bullet$ that can rely on a host of well-measured
modes, i.e. branching ratios that are measured to better
than 10\% accuracy,

$\bullet$ where final states with (multi)neutrals are
included;

$\bullet$ that is performed for $D_s$, $\Lambda _c$ and
$\Xi _c$ two-body channels as well and

$\bullet$ that treats Cabibbo-allowed, Cabibbo-suppressed
and doubly-Cabibbo-suppressed decays separately.

It is quite likely -- and actually I firmly expect --
that such a comprehensive analysis, in particular for the $a_2$
transitions, will reveal serious and systematic
deficiencies in these phenomenological models. To overcome
those will require a description that is more firmly
grounded in QCD. I will briefly comment on them
below. However
I would like first to stress what we have learnt and are
still learning from the phenomenological treatments:

-- They have yielded quite a few successful predictions
in a ``user-friendly'' fashion, in
particular also for $B$ decays.

-- They have helped us significantly to focus on the underlying
theoretical problems, such as the question of factorization or the
$1/N_C$ rule \cite {BURAS}.

-- They can provide us with valuable, albeit indirect, information
on the strong final state interactions. Such information is
crucially important for deciding on the most promising strategy
to search for CP violation in charm decays and then to interpret
properly a signal that might be observed in the future. The so far
most detailled attempt in this direction has recently been undertaken
by the Rome-Napoli group \cite {PUGLIESE}.
The authors conclude that the typical scale for direct
CP asymmetries in D decays is of order $10^{-3}$ -- in agreement
which general expectations stated a few years ago \cite {BIGICHARM}.

\subsection {Theoretical Descriptions}

The first treatment of charm two-body
decays that is intrinsically connected to QCD
was given by Blok and Shifman some time ago, based
on QCD sum rules \cite {BLOKI}. It
would be desirable to have this analysis updated and refined,
in particular by not imposing $SU(3)_{fl}$ symmetry.

The same two authors actually have moved into a different
direction, namely to apply methods based on a heavy quark expansion
to non-leptonic two-body modes of beauty \cite {BLOKII}.
I believe this
ansatz promises to significantly
advance our understanding of exclusive
heavy flavour decays and thus deserves increased attention.

\section{RARE DECAYS}

There are five categories of rare decays that I would like to
sketch.

\subsection {Expected and Informative Decays}

Doubly Cabibbo suppressed decays like
$D^+\rightarrow K^+ \pi ^0$ or
$D^0\rightarrow K^+ \pi ^-$ make up this category. The first
clear evidence for them has been presented recently by the
CLEO collaboration \cite {WITH}:
$$\frac {\Gamma (D^0\rightarrow K^+\pi ^-)}
{\Gamma (D^0\rightarrow K^-\pi ^+)}=(0.77\pm 0.25\pm 0.25)\%
\simeq 3\cdot \tan ^4(\theta _c)\eqno (8)$$ in agreement with
a prediction of $2\cdot \tan ^4(\theta _c)$ for this ratio
\cite {DCSD}.

\subsection {Expected and Unexciting Decays}

The mode $D^0\rightarrow \bar K^{0*}\gamma $ can proceed by
W exchange together with photon bremsstrahlung; a very
rough order of magnitude estimate yields an expected
branching ratio of $\sim {\cal O}(10^{-5})$. Seen by itself
this mode is quite unremarkable. Yet its observation would
serve an ulterior motive. For it has been suggested that the
KM parameter $|V(td)|$ can be extracted from radiative
$B$ decays \cite {ALI}: $BR(B\rightarrow \rho \gamma)/
BR(B\rightarrow K^* \gamma)\simeq |V(td)|^2/|V(ts)|^2$ with
$|V(ts)|\simeq |V(cb)|$. This is based on the assumption that
both radiative transitions are dominated by the
electromagnetic penguin operator. There is however a fly in the
ointment for this interesting suggestion: also W exchange
coupled with photon emission generates
$B\rightarrow \rho \gamma$ transitions; yet this contribution
is independant of $|V(td)|$ and a priori it could be comparable
in size to the penguin-mediated $B\rightarrow \rho \gamma$.
Ignoring such a contribution would lead to the extraction
of an incorrect number for $|V(td)|$. Yet once the
corresponding transition $D\rightarrow \bar K^{0*}\gamma$
has been measured, one can obtain a reliable estimate for the
W exchange contribution to $B\rightarrow \rho \gamma $.

\subsection {Unexpected and Exciting Decays}

Non-minimal SUSY can generate $D^0\rightarrow \rho \gamma$ with
a branching ratio of $\sim 10^{-6}$ - $10^{-5}$
\cite {BGM}. In addition,
Weak Annihilation can also contribute with a branching
ratio $\sim 10^{-6}$. Observing
$$\frac {BR(D^0\rightarrow \rho ^0\gamma )}{BR(D^0
\rightarrow \bar K^{0*}\gamma )}\neq
\tan ^2(\theta _c)$$
would, however, be evidence for the intervention of
``New physics'' like non-minimal SUSY.

\subsection {Puzzling Decays}

An observation of, for instance,
$D^+\rightarrow K^{*+}\gamma $
would be very puzzling. For one expects in the Standard
Model: $BR(D^+\rightarrow K^{*+}\gamma)\sim
\tan ^4(\theta _c)
\cdot BR(D^0\rightarrow \bar K^0\gamma )\sim 10^{-7}$; I
have not found a reasonable New Physics scenario that would
raise this number significantly.

\subsection {``Shots In The Dark''}

I have been unable so far to identify New Physics scenarios
that would generate decays like $D\rightarrow \gamma \gamma ,\;
\gamma \; +\; nothing,\; \pi /\rho \; +\; nothing$ on a level that
could be observed at a $\tau$-charm factory.

\section {SUMMARY AND OUTLOOK}

The defining goal in studies of the physics of charm decays is
to probe and understand the strong interactions in a novel
environment, namely at the interface between the short-distance
and long-distance regime where the effects are numerically large,
but not overwhelming. The prospects of reaching that goal
have been enhanced considerably by the increasing maturity of
second-generation theoretical technologies, namely heavy quark
expansions, QCD sum rules, and QCD simulations on the lattice.
Actually achieving it would be of significant intellectual value;
in addition -- and maybe even more important -- it would
sharpen our theoretical tools for dealing with beauty decays
and fully exploiting the tremendous discovery potential there.
For the non-perturbative corrections that have to be understood with
high precision and reliability in beauty decays are quantitatively
enhanced in charm decays; those act -- as stated before -- as a
microscope.

To match these objectives the experimental program has to be based on
the following elements:

$\bullet$ One has to determine some {\em absolute} branching ratios
precisely, among them semileptonic branching ratios and
lepton spectra.

$\bullet$ A detailed study of the {\em inclusive} semileptonic
$c\rightarrow s$ and $c\rightarrow d$ decays is of particular
value since it prepares us in an optimal way for dealing with
semileptonic $B$ decays at a beauty factory operating at
threshold.

$\bullet$ A {\em large} body of well measured branching ratios --
including for final states containing (multi)neutrals -- has to
be obtained.

$\bullet$ Such a comprehensive program has to be performed
for $D^0$, $D^+$, $D_s$, $\Lambda _c$, $\Xi _c^+$, $\Xi _c^0$,
and preferably also for $\Omega _c$ decays.

A $\tau$-charm factory is optimally suited for this demanding
program and one can thus state that {\bf ``The $\tau$-charm
factory is the QCD machine for the 90's!''}. From the
program listed above it can be concluded that as far as charm physics
is concerned

$\bullet$ one has to cross at least the $\Xi _c \bar \Xi _c$
and preferably the $\Omega _c \bar \Omega _c$ threshold,
i.e. c.m. energies of 5--5.5 GeV have to be reached;

$\bullet$ one needs high luminosity; this is not based on the
outlandish needs of a single measurement, but on the fact that
$D$, $D_s$, $\Lambda _c$ and $\Xi _c$ decay studies cannot be
done in a parasitic fashion, but require separate runs at different
energy settings.

In my judgement it would be inappropriate to invoke searches
for $D^0$--$\bar D^0$ oscillations, rare $D$ decays and CP violation
in charm decays as {\em primary} motivations for building a
$\tau$-charm factory of high luminosity. On the other hand
they represent the ``icing on the cake'': There is a good
chance that such ``High Impact'' physics will emerge in the
charm sector; the high luminosity and purity of the data
sample from a $\tau$-charm factory will be critical in
uncovering such a fundamental surprise.

{\bf Acknowledgements:}
I gratefully acknowledge many illuminating discussions with
my colleagues, in particular with N. Uraltsev, M. Shifman
and B. Blok. I truly enjoyed the stimulating atmosphere created by
the organizers of the Marbella meeting. This work was supported
by the National Science Foundation under grant number
PHY 92-13313.

\end{document}